\documentclass [11pt]{article}
\usepackage{amsmath,amsthm,amsfonts,amscd,eucal,latexsym,amssymb}
\newsymbol\bt 1202           

\newsymbol\rest 1316         

\def\phi{\varphi}

\begin{document}


\hfill{\sl preprint -  }
\par
\bigskip
\par
\rm


\par
\bigskip
\LARGE
\noindent
{\bf }
\bigskip
\par
\rm
\normalsize

\par
\bigskip
\LARGE
\noindent
{\bf Analytic continuation of the Hurwitz Zeta Function
with physical application.}
\bigskip
\par
\rm
\normalsize

\large
\noindent {\bf Vittorio Barone Adesi}

\large
\smallskip

\noindent
Department of Mathematics,
Trento University,\\
I-38050 Povo (TN), Italy.\\
E-mail: adesi@science.unitn.it\\

\large
\noindent {\bf Sergio Zerbini}

\large
\smallskip

\noindent
Department of Physics,
Trento University,\\ 
Gruppo Collegato INFN 
I-38050 Povo (TN), Italy.\\
E-mail: zerbini@science.unitn.it\\

\rm\normalsize



\par
\bigskip
\par
\hfill{\sl September 2001}
\par
\medskip
\par\rm


\noindent
{\bf Abstract:}
A new formula relating the analytic continuation of
the Hurwitz zeta function to the Euler gamma function and a polylogarithmic function is presented. In particular,  the values of the first derivative of the 
real 
part of the analytic continuation of the Hurwitz zeta
function for even negative integers and the imaginary one for odd negative integers are explicitly given.
The result can be of interest both on mathematical and physical side, because 
we are able to apply our new formulas in the context
of the Spectral Zeta Function regularization,  computing the exact pair production rate per space-time unit of
massive Dirac particles interacting with a purely electric background field.

\section{Introduction}

In 1951, Schwinger \cite{Schwinger:1951nm} computed the implicit Effective Lagrangian for a Dirac charged
spinor in general electromagnetic background field using proper time approach.
 In particular, he applied the result to the physically  case of a constant 
and uniform
 electric background and he was able to evaluate 
the exact pair-production rate per space-time unit, namely
\begin{equation}
\label{schw}
w(E,e,m)=\frac{e^{2}E^{2}}{4\pi ^{3}}\sum _{n=1}^{+\infty }\frac{e^{-n\frac{\pi m^{2}}{eE}}}{n^{2}}\:.
\end{equation}

This result is well known and can be obtained by other ways (e.g. Itzykson and
Zuber \cite{2}, using the S-matrix approach, and by Beneventano and Santangelo \cite{3}). In 1990 Blau, Visser and Wipf
\cite{Blau:1991iz}, using the techniques of
the Spectral Zeta Function regularization, tried to obtain the same results 
obtained by
Schwinger. However they were able  
only to obtain non exact results for the four dimensional case,
 using asymptotic analysis. In 1996, Soldati and Sorbo \cite{Soldati:1998vq} obtained a new expression,
but once again their methods were based on asymptotic analysis and the results were expressed in terms of
asymptotic series.

The problem for general electromagnetic external field 
has been recently discussed  by Schubert \cite{Schubert:2000yt} and by 
Cho and Pak \cite{Cho:2001ei}. Schubert obtained the same action by Schwinger  using proper time method and techniques of 
computation inspired by String Theory. Cho and Pak  obtained a 
renormalized action from the general one (non renormalized) 
found by Schwinger, using the so-called  Sitacaramachandrarao identity.
Recently, Beneventano and Santangelo \cite{5} have compared their results 
obtained using Spectral Zeta Function methods  with the 
general result obtained by Schwinger, showing that they agree.

In this paper, we shall compute the imaginary part of the Effective Lagrangian 
related to a Dirac field in a constant and uniform external electric field by
making use of new formulas concerning the analytic continuation of the 
first derivative of the  Hurwitz zeta function.

The content of the paper is the following. In   Section 2, we  briefly 
introduce the Spectral Zeta Function regularization procedure and illustrate 
the result obtained by Blau Visser and Wipf, which will be our 
starting point for the application of the new formulas. In  Section 3,
  we derive a new expression for the analytic continuation of the Hurwitz 
zeta function,
viewed as an analytic function of two complex variables. In Section 4, we
use that result to recover the rate of pair creation of Dirac particles in 
constant and uniform electric
background.

\section{Zeta Function regularization}

In this  Section, we shall review  
the relation between the mathematical problem that we have
solved and the physical problem associated with the  Dirac pair creation.
For a review of  the method, see, for example,
 \cite{7}, \cite{Moretti:1999rf}, \cite{Bytsenko:1996bc} and references quoted therein.

Within the context of Quantum Field Theory (QFT) interacting with a classical gauge background field, one is forced to confront with the determinants of 
differential operators.\\
In the context of a Klein Gordon field interacting with an external gauge field $A$, 
this determinant is related to some physical quantities formally obtained from the $Euclidean$ functional integral:
\begin{equation}
\label{measure}
Z[A_E]=\int D\phi \; e^{-\frac{1}{2} \int d^{4}x \phi A_E \phi},
\end{equation}
where $A_E$ is the Euclidean Klein Gordon operator.
In the definition of this integral we have to analytically continue some global Minkowski temporal coordinate $x^{0}$  into imaginary value $x^{0} \rightarrow ix^{0}$ and consider the analytical continuation of all relevant quantities. \\
The above Gaussian functional integral could be interpreted as a Wiener
 measure, but, 
for our purposes, we may  interpret it in terms of a functional determinant, and rewrite the 
definition of (\ref{measure}) as:
\begin{equation}
Z[A_{E}]=\left[ \mbox{det} \left( \frac{A_{E}}{\mu} \right) \right]^{-\frac{1}{2}} \: .
\end{equation}
where $\mu$ is a constant with the same physical dimension of the operator $A_E$.

It is sometime useful to introduce two physical quantities: the Effective Action and the Effective Lagrangian.
The former is defined  as the logarithm of $Z$. The latter is a function of space-time points
and gives, after an integration on the whole space-time, the Effective Action.
The physical interpretation of this determinant is found after a re-analytical continuation of the imaginary time into a real time.
The result is the vacuum to vacuum transition amplitude.\\ 

We can use Zeta Function regularization to give a rigorous meaning to 
functional  determinants. This regularization techniques was introduced
by Ray-Singer for elliptic differential operators \cite{13}. Within the 
Quantum  Field Theory, it was used by by 
 Dowker and Chritchley \cite{Dowker:1977zf}  and Hawking
\cite{Hawking:1977ja}. Given a  compact Riemannian manifold $M$ and for  elliptic and second order operators acting on $L_2(M)$,  it can be proved that such
 definition gives a useful extension of the notion of functional determinant.

 Since the square of the Euclidean Dirac operator is a elliptic second order 
 differential operator, making  use of the zeta regularization 
technique, Blau, Visser and Wipf \cite{Blau:1991iz} arrived at the following 
Effective Lagrangian for a Dirac field in an external constant and uniform electric field:
\begin{equation}
\label{leff}
L_{eff}(E,0)=-\frac{e^{2}E^{2}}{2\pi ^{2}}\{[1-\mbox{ln}
(\frac{-2ieE}{\mu ^{2}})]\zeta _{H}(-1;1+i\frac{m^{2}}{2eE})+
\end{equation}
\begin{equation}\nonumber
+\frac{d}{ds}\zeta _{H}(s;1+i\frac{m^{2}}{2eE})_{s=-1}\}+i\frac{m^{2}eE}{8\pi ^{2}}[\mbox{ln}\frac{m^{2}}{\mu ^{2}}-1]\, .
\end{equation}
  where $\zeta_{H}(s;x)$ is the Hurwitz zeta function defined by:
\begin{equation}\nonumber
\zeta_{H}(s;x):=\sum^{\infty}_{n=0}\left(\frac{1}{n+x}\right)^s\:,\,\,\,\,
\mbox{Re}s>1 \,.
\end{equation}
   
They were not able to find an explicit form for the analytic continuation of 
the derivatives of the Hurwitz zeta function at $s=-1$.  Thus, a 
direct comparison between their result and the one  obtained by Schwinger 
was missing.

\section{Analytic continuation of the Hurwitz zeta function}

In this Section, we discuss the analytical properties of the Hurwitz zeta 
function.
Following \cite{8}, we obtain a new identity involving 
the Hurwitz zeta function, the Euler $\Gamma(s)$ function and a polylogarithmic function. 

Recall that the  Hurwitz zeta function is  defined as follows:
\begin{equation}
\label{Hurwitz}
\zeta_H(s;x):=\sum _{n=0}^{+\infty }(n+x)^{-s},
\end{equation}
where $\mbox{Re s} >1$, $ x \neq 0, -1, -2,... $.
One can analytically extend it into an analytic function of two complex 
variables  $s$ and $x$.

In order to search for the analytical continuation in the double complex plane, we introduce the function $F(s;z)$
defined by the following series:
\begin{equation}
\label{zeta}
F(s;z):=\lim_{k\rightarrow\infty}\sum_{n=-k}^{k}(-in+z)^{-s}\, .
\end{equation}
Notice that the  series above is absolutely convergent  for  $\mbox{Re s}> 1$,
$z \neq \pm ik, k \in N$. 
Thus, one may write:
\begin{equation}
\label{decomp}
F(s;z):=\lim_{k\rightarrow\infty}\left[ \sum_{n=0}^{k}(-in+z)^{-s}+\sum_{n=0}^{k}(in+z)^{-s}\right]-z^{-s}.
\end{equation}
\\
Using this definition for the series (\ref{zeta}), we proceed in searching for its analytical continuation.
With regard to this issue, previous attempts  can be found in \cite{Tomszeta}, \cite{Toms:1995rr}. Here the author obtained an expression different from our one, requiring, however, as far as   
physical  applications were concerned, an ad hoc finite renormalization.

It is straightforward to prove, the following relation between $F(s;z)$  and 
$\zeta_H(s;z)$:
\begin{equation}
\label{relation}
F(s;z)=i^{s}\zeta _{H}(s;iz)+i^{-s}\zeta _{H}(s;-iz)-z^{-s}.
\end{equation}

First, let us investigate the  function $F(s;x)$.
The following theorem gives an expression for the analytic continuation of $F(s;z)$ in terms of
polylogarithmic function.
\\
\\
{\bf Theorem:}
Following the definition stated above, the analytic continuation of $F(s,z)$ for each $s\in C$ and
$\mbox{Re}$ $z>$ 0 is given by\\
\begin{equation}
\label{master}
F(s;z)=\frac{(2\pi )^{s}}{\Gamma (s)} \sum _{n=1}^{+\infty }\frac{e^{-2n\pi z}}{n^{1-s}}.
\end{equation}\\
Proof:
Define the sequence $F_k (s;z):$
\begin{equation}
F_k (s;x):=\sum_{n=-k}^{k}(-in+z)^{-s}, 
\end{equation}
which obviously converges to $F(s;z)$.\\
Recall the definition of the Euler $\Gamma(s)$ function for $\mbox{Re s}>0$,
 \begin{equation}\nonumber
\Gamma (s):=\int _{0}^{\infty }t^{s-1}e^{-t}dt\, .
\end{equation}
Consider the product of this function and the sequence $F_k (s;x)$ :
\begin{equation}\nonumber
F_k (s;z)\Gamma(s) = \sum_{n=-k}^{+k }(-in+z)^{-s}\int _{0}^{\infty}t^{s-1}e^{-t}dt=
\int _{0}^{\infty}\sum_{n=-k}^{+k }(-in+z)^{-s}t^{s-1}e^{-t}dt\: .
\end{equation}
We may change variables according to $t=\tau(-in+z)$, which is allowed for $z$ not purely imaginary integer since 
$|\frac{dt}{d\tau}|\neq0$, and we obtain:

\begin{equation}
\label{distrprod}
F_k(s;z)=\frac{1}{\Gamma (s)}\int ^{+\infty }_{0}\,d\tau
e^{-\tau z}{\tau}^{s-1}\sum _{n=-k }^{+k }e^{in\tau}\,.
\end{equation}

The part of the integrand in (\ref{distrprod}) multiplying the series 
 $vanishes$ in $\tau=0$ for  $\mbox{Re s} > 1$. This implies that 
the integration may be interpreted as the action of the k-element of a sequence of distributions
$g_k(\tau)$,
\begin{equation}
g_k(\tau):=\sum _{n=-k }^{+k }e^{in\tau}
\end{equation}
on the continuous function $e^{-\tau z}{\tau}^{s-1}$.
It is a well-known result
 of the theory of distribution that this sequence converges and gives the Poisson summation formula:
\begin{equation}\nonumber
\lim_{k\rightarrow\infty}g_k(\tau) = \lim_{k\rightarrow\infty}\sum _{n=-k }^{+k }e^{in\tau} = 2\pi \sum _{n=-\infty }^{+\infty }\delta
(\tau-2n\pi ).
\end{equation}

Thus we obtain in the limit $k\rightarrow\infty$:
\begin{equation}\nonumber
F(s;z)=\frac{1}{\Gamma (s)}\int ^{+\infty }_{0 }e^{-\tau z}{\tau}^{s-1}\sum _{n=-\infty }^{+\infty }e^{in\tau}=
\end{equation}
$$
=\frac{(2\pi)^{s}}{\Gamma (s)}
\sum _{n=1}^{+\infty }\frac{e^{-2n\pi z}}{n^{1-s}},
$$
which, as stated in the hypotheses of the theorem, makes sense as analytic
continuation for each $s\in C$ and Re z $>$0.\\
\\
Recalling the relation (\ref{relation}) between $F(s;z)$ and $\zeta_H(s;z)$, we have the following corollary:\\
\\
{\bf Corollary.} 
The following formula is valid for each $s\in$ $C$ and Re z $>$ 0 in the sense of the analytic continuation,\\
\begin{equation}
\label{facile}
(i)^{s}\zeta _{H}(s;iz)+i^{-s}\zeta _{H}(s;-iz)-z^{-s}=\frac{(2\pi )^{s}}{\Gamma (s)}\sum _{n=1}^{+\infty}\frac{e^{-2n\pi{z}}}{n^{1-s}}.
\end{equation}
Proof: Direct use of equations (\ref{relation}) and (\ref{master}).
\\
\\
Equation (\ref{facile}) can be checked for $s=-n$. In this case the right hand side is vanishing and
equation (\ref{facile}) gives:
\begin{equation}
{i}^{-n}\zeta_H(-n,iz) +i^{n}\zeta_H(-n,-iz) - z^{-s} = 0\:.
\end{equation}
Recall that the values of the Hurwitz zeta function are known for negative integers and are related to the
Bernoulli's polynomials by the following
 formula,
\begin{equation}
\zeta_H(-n,z)=-\frac{B_{n+1}(z)}{n+1}.
\end{equation}
It is easy to show that this equation is identically fulfilled.\\
\\
{\bf Remark.} 
We now notice that the series:\\
\begin{equation}
\sum _{n=1}^{+\infty }\frac{e^{-2n\pi z}}{n^{1-s}}
\end{equation}
and its derivatives are uniformly convergent in the variable s for $\mbox{Re z} \geq 0$ and Re s $\in (-\infty,a)$ with $a < \infty$. 
A straightforward computation leads to the following lemma:\\
\\
{\bf Lemma 1.}
The following formula holds for each s $\in$ C and Re z $>$ 0 
in the sense of the analytic continuation,
\begin{equation}\nonumber
(i)^{s}\frac{d}{ds}\zeta_H(s;iz)+ (i)^{-s}\frac{d}{ds}\zeta_H(s;-iz) = 
(2\pi)^s (\frac{d}{ds}\frac{1}{\Gamma(s)})\sum _{n=1}^{+\infty}\frac{e^{-2n\pi{z}}}{n^{1-s}} + 
\frac{(2\pi)^{s}}{\Gamma(s)} \sum _{n=1}^{+\infty}
\frac{e^{-2n\pi{z}}\mbox{ln}(2n\pi)}{n^{1-s}}+
\end{equation}
\begin{equation}
+i\frac{\pi}{2}\lbrack{ -2(i)^{s}\zeta_H(s;iz)+z^{-s} + \frac{(2\pi)^s}{\Gamma(s)}\sum
_{n=1}^{+\infty}\frac{e^{-2n\pi{z}}}{n^{1-s}}}\rbrack -
(z)^{-s}(\mbox{ln}|z| + iarg\:{z})\: . 
\end{equation}
Proof: 
One simply applies the derivative on both sides of the formula obtained in the corollary above, then uses
the remark above to exchange the series with the derivative with respect to $s$,
 and then analytically continues the result for all the values of $s$.\\
\\

As a result,  we shall derive formulas for the real and imaginary values of 
the first derivative of the Hurwitz zeta 
function respectively for even and odd 
negative integers.\\
Using parity properties of the real and the imaginary part of this polynomials it is easy to prove that,  \\
\\
{\bf Lemma 2.}
For $x\in R$, $x>0$, and for $n\in N$, $n<0$, the Hurwitz zeta function has the following parity properties:
\begin{equation}
\mbox{Re} \: \zeta_H(-n;ix) = \mbox{Re} \:\zeta_H(-n;-ix)
\end{equation}
\begin{equation}
\mbox{Im} \: \zeta_H(-n;ix) = -\mbox{Im} \:\zeta_H(-n;-ix)
 \end{equation} 
 \\
 \\
We have
\\
\\
 {\bf Proposition.}
  For $x\in R$ and for $m\in N$, the following formulas two are valid for any natural number:
  \begin{equation}
  \label{odd}
 \mbox {Im} \frac{d}{ds} \: \zeta_H(-(2m+1);ix)=\pi\: \mbox{Re}\: \frac{{B_{2m+2}(ix)}}{4(m+1)}+\frac{(-1)^{m+1}x^{2m+1}}{2}\mbox{ln} \:x
 +\frac{(-1)^{m+1}(2m+1)!}{2(2\pi)^{2m+1}}\sum_{n=1}^{+\infty}\frac{e^{-2n\pi{x}}}{n^{2(m+1)}}
\end{equation}
\begin{equation}
\label{even}
 \mbox Re\: \frac{d}{ds}\zeta_H(-2m;ix)=-\pi\: \mbox{Im}\: \frac{{B_{2m+1}(ix)}}{2(2m+1)}+\frac{(-1)^{m+1}x^{2m}}{2}\mbox{ln} \:x
 +\frac{(-1)^{m}(2m)!}{2(2\pi)^{2m}}\sum_{n=1}^{+\infty}\frac{e^{-2n\pi{x}}}{n^{1+2m}}
 \end{equation}
 Proof: Simple computation using Lemma 1 and 2.\\
 \\
\\
{\bf Note.}
It is possible to obtain other identities involving higher order derivatives 
 of the Hurwitz zeta function starting from (\ref {facile}).
As a consequence, it is possible to obtain the values
of higher derivatives of either the imaginary or real part of the Hurwitz zeta 
function for negative integers using recursive formulas.

As far as we know only asymptotic values have been found by E.Elizalde
\cite{Elizalde:1993zg,19} for the analytic continuation of the 
 derivatives of the Hurwitz zeta function. Thus (\ref{odd}) and (\ref{even}) 
are new formulas. In the next Section, we use the second one for the application to 
the study of the Dirac pair creation in pure electrical background field.
\\
\\
\section{Application to Schwinger pair creation}

In this last Section we shall apply the formula (\ref{odd}) to the problem of 
Dirac pair creation in a purely electrical background field,
recovering the Schwinger's result.
The starting point is  the effective Lagrangian obtained by Blau, Visser and Wipf,
\begin{equation}\nonumber
\label{popona} L_{eff}(E,0)=-\frac{e^{2}E^{2}}{2\pi ^{2}}\{[1-\mbox{ln}(\frac{-2ieE}{\mu ^{2}})]\zeta _{H}(-1;1+i\frac{m^{2}}{2eE})+
\end{equation}
\begin{equation}
\frac{d}{ds}\zeta _{H}(s;1+i\frac{m^{2}}{2eE})_{s=-1}\}+i\frac{m^{2}eE}{8\pi ^{2}}[\mbox{ln}\frac{m^{2}}{\mu ^{2}}-1] .
\end{equation}
We have to find the imaginary part of the right-hand side of (\ref{popona}).
Making use of Eq.
(\ref{odd}) to $s=-1$, we get:
\begin{equation}
\mbox Im \: \frac{d}{ds}\zeta_H(-1;ix)= -\frac{1}{4\pi}\sum_{n=1}^{+\infty}\frac{e^{-2n \pi x}}{n^{2}}
-\frac{x}{2}\mbox{ln} \:x
+\frac{\pi}{24}-\frac{\pi x^2}{4}\:.
\end{equation}
If we plug this formula in the effective Lagrangian expression, 
we obtain the Schwinger's result:
\begin{equation}
w(E,e,m)=\frac{e^{2}E^{2}}{4\pi ^{3}}\sum _{n=1}^{+\infty }\frac{e^{-n\frac{\pi m^{2}}{eE}}}{n^{2}}.
\end{equation}

\section{Acknowledgments}
We are grateful to Valter Moretti and Luciano Vanzo (University of Trento) and in  particular to 
E. M. Santangelo and C. G. Beneventano(Department of Physics, Faculty of Ciencias Exactas, National
University of La Plata) for very useful discussion.
\\
\\

\end{document}